\documentclass[aps,amsmath,amssymb,twocolumn,showpacs,pra]{revtex4-1}
\usepackage{graphicx}
\begin{document}

\title{Quantum probabilities from combination of Zurek's envariance and Gleason's theorem}

\author{A.~V.~Nenashev}
\email {nenashev@isp.nsc.ru}
\affiliation{Rzhanov Institute of Semiconductor Physics, 630090 Novosibirsk, Russia}
\affiliation{Novosibirsk State University, 630090 Novosibirsk, Russia}

\date{\today}

\begin{abstract}
The quantum-mechanical rule for probabilities, in its most general form of 
positive-operator valued measure (POVM), is shown 
to be a consequence of the environment-assisted invariance (envariance) idea 
suggested by Zurek [Phys. Rev. Lett. {\bf90}, 120404 (2003)], being completed 
by Gleason's theorem. This provides also a method for derivation of the Born rule.
\end{abstract}

\pacs{03.65.Ta}

\maketitle

\section{Introduction} \label{introduction}

Almost all textbooks on quantum mechanics consider only measurements of a special 
kind---namely, measurements of \emph{observables}. An observable $\mathcal{O}$ corresponds to 
some Hermitian operator $\hat O$. A measuring device measures the observable $\mathcal{O}$, if 
(i) the only possible results of measurements are eigenvalues of $\hat O$, and (ii) if the 
state $|\psi\rangle$ of the system under measurement is an eigenstate  of $\hat O$, one can 
predict the measurement result \emph{with certainty} to be the eigenvalue of $\hat O$ 
corresponding to $|\psi\rangle$. When $|\psi\rangle$ is not an eigenvector of $\hat O$, the 
measurement result cannot be known in advance, but the postulates of quantum mechanics allow one to predict the 
\emph{probabilities} of results. Namely, the probability $p_\lambda\,(|\psi\rangle)$ of the result $\lambda$ is  
\begin{equation} \label{eq:projective}
  p_\lambda(|\psi\rangle) = \langle\psi | \hat P_\lambda | \psi\rangle,
\end{equation}
where $\hat P_\lambda$ is the projector onto the eigenspace of $\hat O$ corresponding to the eigenvalue $\lambda$.
In the simplest case of non-degenerate eigenvalue $\lambda$, the projector $\hat P_\lambda$ is equal to 
$|\varphi_\lambda\rangle\langle\varphi_\lambda|$, 
where $|\varphi_\lambda\rangle$ is the eigenvector, and Eq.~(\ref{eq:projective}) turns into the Born rule:
\begin{equation} \label{eq:born-rule}
  p_\lambda(|\psi\rangle) = 
  \langle\psi | \varphi_\lambda\rangle\langle\varphi_\lambda | \psi\rangle
  \equiv \bigl| \langle\varphi_\lambda|\psi\rangle \bigr|^2 .
\end{equation}

Unlike this special class of measurements, \emph{general measurements} are not associated with any observables, 
and probabilities of their results do not obey literally Eq.~(\ref{eq:projective}). Instead, the probability 
$p_\lambda(|\psi\rangle)$ of some result $\lambda$ of a general measurement can be expressed as 
\begin{equation} \label{eq:povm}
  p_\lambda(|\psi\rangle) = \langle\psi | \hat A_\lambda | \psi\rangle ,
\end{equation}
where $\hat A_\lambda$ is some Hermitian operator (not necessary a projector). 
The set of operators $\{\hat A_\lambda\}$ obeys the following requirements, which are consequences of 
properties of probability:
\newline (1) eigenvalues of operators $\hat A_\lambda$ are bound within the range [0,1];
\newline (2) the sum $\sum_\lambda \hat A_\lambda$ (over all measurement results $\lambda$) is equal to the 
identity operator.
\newline The set $\{\hat A_\lambda\}$ satisfying these requirements is usually called a \emph{positive-operator 
valued measure} (POVM)~\cite{Peres,Nielsen}.

Such general measurements, described by POVMs via Eq.~(\ref{eq:povm}), occur in various contexts: 
as \emph{indirect measurements}, when a system~$A$ (to be measured) first interacts with another quantum 
system~$B$, and actual measurement is then performed on the 
system~$B$~\cite{Peres,Nielsen,Braginsky}; 
as \emph{imperfect measurements}, where a result of a measurement is subjected to a
random error~\cite{Braginsky,Gardiner}; 
as \emph{continuous} and \emph{weak measurements}~\cite{Jacobs2006}, etc.

In this paper, we will show that the rule~(\ref{eq:povm}) for probabilities of results of general measurements 
is a simple consequence of Gleason's theorem. This theorem~\cite{Gleason1957,Dvurecenskij} is a key statement 
for quantum logics, and also can be considered as a justification of the Born rule~\cite{Dickson2011}. But the 
usual way of getting the probability rule from Gleason's theorem requires \emph{non-contextuality} to be 
postulated~\cite{Peres,Dickson2011}. We will show that it is possible to avoid the demand of non-contextuality.

We will use Gleason's theorem in the following (somewhat restricted) formulation. Let $p\,(|e\rangle)$ be 
a real-valued function of unit vectors $|e\rangle$ in $N$-dimensional Hilbert space. Suppose that
\newline (1) $N \geq 3$,
\newline (2) the function $p$ is non-negative,
\newline (3) the value of the sum
\begin{equation} \label{eq:gleason-condition}
  \sum_{n=1}^N p\,(|e_n\rangle),
\end{equation}
where unit vectors $|e_1\rangle, |e_2\rangle, \ldots |e_N\rangle$ are all mutually orthogonal, 
does not depend on the choice of the unit vectors.
\newline Then, Gleason's theorem states that the function $p\,(|e\rangle)$ can be represented as follows:
\begin{equation} \label{eq:gleason-answer}
  p\,(|e\rangle) = \langle e | \hat A | e \rangle ,
\end{equation}
where $\hat A$ is some Hermitian operator in the $N$-dimensional Hilbert space.

We will apply Gleason's theorem in a quite unusual way. Typically, the argument $|e\rangle$ is considered 
as a property of a measuring device, and the function $p$ as a characteristic of the measured system's state. 
Our approach is completely reverse---we interpret the vector $|e\rangle$ as a system's state vector, 
and refer the function $p$ to a measuring device. The main difficulty of this approach consists in satisfying 
the third condition: namely, that the sum~(\ref{eq:gleason-condition}) is constant. 
To show that this condition fulfils, we will exploit the concept 
of environment-induced invariance, or \emph{envariance}, suggested by Zurek~\cite{Zurek2003,Zurek2005}. The idea of 
\emph{envariance} can be formulated as follows: when two quantum systems are entangled, 
one can \emph{undo} some actions with the first system, performing corresponding “counter-actions” with the second one. 
Such a possibility of undoing means that these actions do not change the state of the first system (considered as alone) 
and, in particular, do not change probabilities of results of any measurements on this system~\cite{Zurek2003,Zurek2005}. 
Note that envariance was introduced by Zurek as a tool for understanding the nature of quantum probabilities, 
and for derivation the Born rule.

For illustrative purposes, we will depict a quantum system as a moving particle, and a measuring device---as 
a black box (that emphasizes our ignorance about construction of this device and about processes inside it), see Fig.~1.
When the particle reaches the black box, the lamp on the box either flashes for a moment, or stays dark. 
One can introduce the probability $p\,(|\psi\rangle)$ of flashing the lamp, as the rate of flashes normalized to 
the rate of particle arrivals, when all these particles are in the state $|\psi\rangle$. The main result of the 
present paper consists in finding out that 
\begin{equation} \label{eq:answer}
  p\,(|\psi\rangle) = \langle \psi | \hat A | \psi \rangle ,
\end{equation}
with some Hermitian operator $\hat A$.
\begin{figure}
\includegraphics[width=0.72\linewidth]{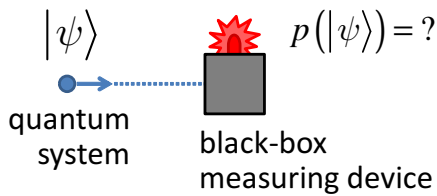} 
\caption{Probability $p\,(|\psi\rangle)$ of flashing the lamp as a result 
of interaction of the quantum system in a state $|\psi\rangle$ 
with the measuring device.}
\label{fig1}
\end{figure}

Though we consider a measurement with only two possible results (flashing and non-flashing of the lamp), 
this does not lead to any loss of generality. Indeed, one can associate flashing of the lamp with some particular 
measurement result $\lambda$, and non-flashing---with all other results. Then, the function $p\,(|\psi\rangle)$ 
in Eq.~(\ref{eq:answer}) would be the same as the function $p_\lambda(|\psi\rangle)$ in Eq.~(\ref{eq:povm}). 
So any proof of Eq.~(\ref{eq:answer}) also proves Eq.~(\ref{eq:povm}), i.~e. justifies the POVM nature of 
every conceivable measurement.

For simplicity, we restrict ourselves by consideration of quantum systems with \emph{finite-dimensional} state spaces.

In Section~\ref{envariance} we will introduce a particular case of \emph{envariance}, which will be used later. 
Section~\ref{preparation} illustrates preparation of a quantum system in a pure state by measurement of \emph{another} 
system. In Section~\ref{thought-exp}, we will consider a series of thought experiments that 
combine the features discussed in previous two sections. These experiments show that the function 
$p\,(|\psi\rangle)$ obeys Eq.~(\ref{eq:identity1}). In Section~\ref{gleason-appl} we will demonstrate that 
Eq.~(\ref{eq:identity1}) together with Gleason's theorem lead to the probability rule~(\ref{eq:answer}). 
In Section~\ref{2d}, the special case of two-dimensional state space (not covered directly by Gleason's 
theorem) is considered. Finally, Section~\ref{born-rule} shows how the projective postulate~(\ref{eq:projective}) 
(and the Born rule as a particular case) follows from the POVM probability 
rule~(\ref{eq:answer}). Closing remarks are gathered in Section~\ref{conclusions}.

\section{Envariance} \label{envariance}

\begin{figure} 
\includegraphics[width=0.86\linewidth]{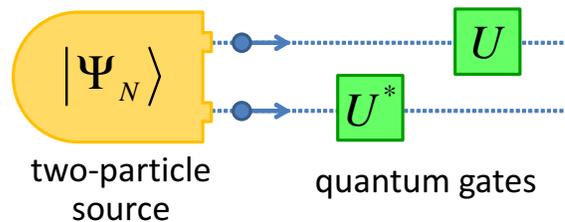}
\caption{Illustration of \emph{envariance}: the initial joint state of two particles, perturbed by the gate 
$U^*$, is restored after applying the gate $U$ to another particle.}
\label{fig2}
\end{figure}

Let us consider an experiment shown in Fig.~\ref{fig2}. Two identical particles are prepared in the joint state 
\begin{equation} \label{eq:psi-N}
  |\Psi_N\rangle = \frac{|1\rangle|1\rangle + |2\rangle|2\rangle +\ldots+ |N\rangle|N\rangle }{\sqrt N} \,,
\end{equation}
$|1\rangle, |2\rangle, \ldots, |N\rangle$ being some orthonormal basis of the $N$-dimensional state space 
of one particle. 
After that, each particle passes through a \emph{quantum gate}, i.~e. a device that performs some unitary 
transformation under the corresponding particle. For the first (upper) particle, an arbitrarily chosen 
transformation $\hat U$ is used. For the second (lower) particle, the complex-conjugated transformation 
$\hat U^*$ (whose matrix elements in the basis $|1\rangle, \ldots, |N\rangle$ are complex conjugates to 
corresponding matrix elements of $\hat U$) is applied.

Let us find the joint state $|\Psi'_N\rangle$ of two particles after passing through the gates:
\begin{equation} \label{eq:transform}
  |\Psi'_N\rangle = \frac{1}{\sqrt N} 
  \sum_{n=1}^N \left(\hat U|n\rangle\right) \left(\hat U^*|n\rangle\right) ,
\end{equation}
where 
\begin{equation} \label{eq:u-n}
  \hat U|n\rangle = \sum_{k=1}^N U_{kn}|k\rangle 
\end{equation}
($U_{kn}$ being matrix elements of $\hat U$), and 
\begin{equation} \label{eq:u-star-n}
  \hat U^*|n\rangle = \sum_{l=1}^N (U_{ln})^*|l\rangle .
\end{equation}
Substituting the latter two equalities into Eq.~(\ref{eq:transform}), and changing the order of summation, 
one can get
\begin{equation} \label{eq:transform2}
  |\Psi'_N\rangle = \frac{1}{\sqrt N} 
  \sum_{k=1}^N \sum_{l=1}^N \left( \sum_{n=1}^N U_{kn} (U_{ln})^* \right) |k\rangle|l\rangle .
\end{equation}
Due to unitarity of the matrix $U$, the expression in brackets in Eq.~(\ref{eq:transform2}) reduces to the 
Kroneker's delta $\delta_{kl}$:
\begin{equation} \label{eq:unitarity}
  \sum_{n=1}^N U_{kn} (U_{ln})^* = \delta_{kl} .
\end{equation}
Hence,
\begin{equation} \label{eq:psi-pp-is-psi}
  |\Psi'_N\rangle = \frac{1}{\sqrt N}   \sum_{k=1}^N |k\rangle|k\rangle  \equiv  |\Psi_N\rangle .
\end{equation}

Thus, effects of two transformations $\hat U$ and $\hat U^*$, applied to different entangled particles 
prepared in the joint state $|\Psi_N\rangle$, Eq.~(\ref{eq:psi-N}), cancel each other. 
According to Zurek~\cite{Zurek2003,Zurek2005}, this means that each of these transformations do not change 
the state of the particle, on which it acts. In other words, the state of each particle is invariant 
(``\emph{envariant}'') under such transformations.

The fact that the two-particle state $|\Psi_N\rangle$ remains unchanged when the particles pass through 
the gates $U$ and $U^*$ (Fig.~\ref{fig2}) will be used in Section~\ref{thought-exp}.

\section{Preparation by measurement}  \label{preparation}

Let us consider a special measurement device (a ``meter'') that distinguishes the basis states 
$|1\rangle, |2\rangle, \ldots, |N\rangle$ from each other. Therefore the following property is satisfied by definition:

{\bf Property a.}  If a measured system was in the state $|k\rangle$ before measurement by the meter 
($k\in\{1,2,\ldots,N\}$), then the measurement result will be $k$ with certainty.

It is commonly accepted that any such ``meter'' has to obey also the following property, which is the reversal of 
Property a:

{\bf Property b.}  The only state, for which the result of measurement by the meter can be predicted 
to be $k$ with certainty, is the pure state $|k\rangle$.

\begin{figure}
\includegraphics[width=0.86\linewidth]{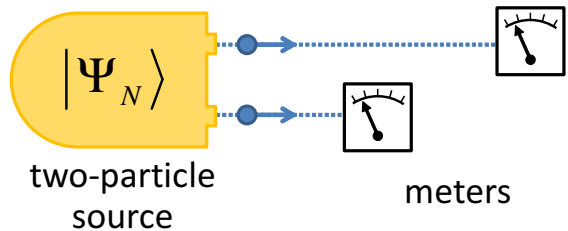}
\caption{After the measurement of the lower particle, the upper one appears in the state $|n\rangle$, 
where $n$ is the measurement result.}
\label{fig3}
\end{figure}

Quantum mechanics also guarantees that the following statement is true:

{\bf Property c.}  If two systems were in the joint state $|\Psi_N\rangle$, Eq.~(\ref{eq:psi-N}), 
and each of them was measured 
by a meter as shown in Fig.~\ref{fig3}, then the results of these two measurements must coincide.

Now consider a state of the upper particle in Fig.~\ref{fig3} just after the lower particle was measured. 
Let $n$ be the measurement result obtained by the lower meter. Then, according to Property c, one can predict that 
the result of the upper particle's measurement will also be $n$. Due to Property b, this means that the upper particle 
is now in the pure state $|n\rangle$.

Hence, if a system of two particles was initially in the state $|\Psi_N\rangle$, and one particle is measured by a 
meter, this measurement \emph{prepares} the other particle in the state $|n\rangle$, where $n$ is the result of 
the measurement. This conclusion will be used in the next Section.

\section{Three thought experiments}  \label{thought-exp}

\begin{figure} 
\includegraphics[width=\linewidth]{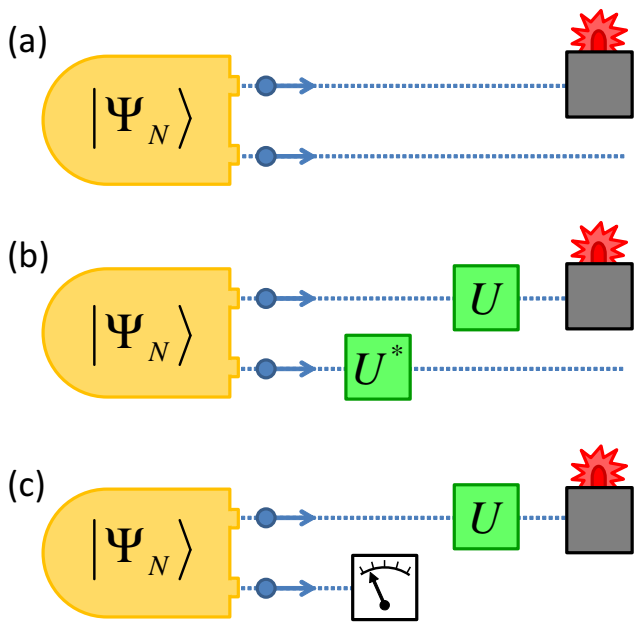}
\caption{Three thought experiments. The probability $\mathcal{P}$ of flashing the light on the 
measuring device is the same in all three experiments.}
\label{fig4}
\end{figure}

Let us examine the measuring device, schematically represented in Fig.~\ref{fig1}, by means of the 
equipment introduced in Figs.~\ref{fig2} and~\ref{fig3}. 
Figure~\ref{fig4}a shows the experiment, in which the source, emitting pairs of particles prepared 
in the state $|\Psi_N\rangle$, is combined with the measuring device. 
One can define the probability $\mathcal{P}$ of flashing the lamp on the device, as a ratio of 
the rate of flashing to the rate of emitting the particles by the source.

In the next thought experiment, Fig.~\ref{fig4}b, two quantum gates $U$ and $U^*$, 
the same as in Fig.~\ref{fig2}, are added on the way of particles. 
It was shown in Section~\ref{envariance} that this combination of 
gates leaves the state $|\Psi_N\rangle$ unchanged. Thus, from the point of view of the measuring device, 
nothing was changed when the two gates were introduced, consequently the rate of flashing of the lamp 
remains unchanged. Thus, the probability $\mathcal{P}$ of flashing the lamp on the measuring device 
is the same in Figs.~\ref{fig4}a and~\ref{fig4}b.

The third thought experiment in this series (Fig.~\ref{fig4}c) differs from the second one (Fig.~\ref{fig4}b) 
by removing the gate $U^*$ and inserting the ``meter'', which measures the state of the lower particle 
in the basis $|1\rangle, |2\rangle, \ldots, |N\rangle$, as in Fig.~\ref{fig3}. 
Since the difference between Fig.~\ref{fig4}b and Fig.~\ref{fig4}c is related to the lower branch of the 
experimental setup only, it cannot influence any events of the higher branch. (Otherwise, it would be 
possible to transfer information from the lower branch to the higher one, without any physical interaction 
between the branches.) So we conclude that the probability $\mathcal{P}$ of flashing the lamp on the 
measuring device in the third experiment is the same as in the second one. 

Now we will express the value of $\mathcal{P}$ in the third experiment (Fig.~\ref{fig4}c) through the 
function $p\,(|\psi\rangle)$ defined in Section~\ref{introduction} (a probability of lamp flashing for 
the pure state $|\psi\rangle$ of the measured particle). 
Let $a_n$ denote the probability that the meter at the lower branch gives the result $n$. 
Also, let $\mathcal{P}_n$ denote the probability that this meter gives the result $n$ 
\emph{and} the lamp on the measuring device at the higher branch flashes. Obviously,
\begin{equation} \label{eq:sum-a}
  \sum_{n=1}^N a_n = 1 ,
\end{equation}
\begin{equation} \label{eq:total-probability}
  \sum_{n=1}^N \mathcal{P}_n = \mathcal{P} .
\end{equation}
If the lower meter gives the result $n$, then the upper particle appears in the state $|n\rangle$, 
according to discussion in Section~\ref{preparation}. After passing through the gate $U$, 
the upper particle's state turns into $\hat U|n\rangle$. Thus, the (conditional) probability of lamp flashing 
on the device at the higher branch is equal to $p\left(\hat U|n\rangle\right)$ \emph{if} 
the lower meter's result is $n$. Then, according to the multiplicative rule for probabilities,
\begin{equation} \label{eq:experiment4}
  \mathcal{P}_n = a_n \, p\left(\hat U|n\rangle\right) .
\end{equation}
Combination of Eqs.~(\ref{eq:total-probability}) and (\ref{eq:experiment4}) gives
\begin{equation} \label{eq:identity1}
  \sum_{n=1}^N a_n \, p\left(\hat U|n\rangle\right) = \mathcal{P} ,
\end{equation}
which is simply a manifestation of the law of total probability applied to the experiment shown in Fig.~\ref{fig4}c.
In Eq.~(\ref{eq:identity1}), the value of $\mathcal{P}$ does not depend on choice of the unitary operator $\hat U$, 
because this value is the same as in the first experiment (Fig.~\ref{fig4}a), see discussion above. 
Also the values of $a_n$ do not depend on $\hat U$.

In the next Section, we will derive Eq.~(\ref{eq:answer}) from Eq.~(\ref{eq:identity1}).

\section{Applying Gleason's theorem}  \label{gleason-appl}

Let $\mathcal{E} = \{ |e_1\rangle, |e_2\rangle, \ldots, |e_N\rangle \}$ be an orthonormal set of vectors 
in the $N$-dimensional Hilbert space: $\langle e_m | e_n \rangle = \delta_{mn}$. Then, it is possible to 
construct an unitary operator $\hat U$ that transforms the set of basis vectors 
$\{ |1\rangle, ..., |N\rangle \}$ into $\mathcal{E}$:
\begin{equation} \label{eq:u-construction}
  \hat U|n\rangle = |e_n\rangle , \quad n = 1, \ldots, N .
\end{equation}
Any such unitary operator can be implemented (at least in a thought experiments) as a physical device (quantum gate). 
Thus, Eq.~(\ref{eq:identity1}) is valid for the operator $\hat U$ defined by Eq.~(\ref{eq:u-construction}). 
Substituting Eq.~(\ref{eq:u-construction}) into Eq.~(\ref{eq:identity1}), one can see that
\begin{equation} \label{eq:identity2}
  \forall \; \mathcal{E}: \quad
  \sum_{n=1}^N a_n \, p\left(|e_n\rangle\right) = \mathcal{P} ,
\end{equation}
where values of $a_n$ and $\mathcal{P}$ do not depend on the choice of $\mathcal{E}$.

Now we will see how to get rid of the unknown coefficients $a_n$. Let us first examine the simplest case of $N=2$. 
Eq.~(\ref{eq:identity2}) for $N=2$ reads:
\begin{equation} \label{eq:identity2-2d-1}
  a_1 \, p\left(|e_1\rangle\right) + a_2 \, p\left(|e_2\rangle\right) = \mathcal{P} .
\end{equation}
If $\{ |e_1\rangle, |e_2\rangle \}$ is an orthonormal set, then, obviously, 
$\{ |e_2\rangle, |e_1\rangle \}$ is also an orthonormal set. Therefore Eq.~(\ref{eq:identity2-2d-1}) 
remains valid if one swaps the vectors $|e_1\rangle$ and $|e_2\rangle$:
\begin{equation} \label{eq:identity2-2d-2}
  a_1 \, p\left(|e_2\rangle\right) + a_2 \, p\left(|e_1\rangle\right) = \mathcal{P} .
\end{equation}
Summing up Eqs.~(\ref{eq:identity2-2d-1}) and~(\ref{eq:identity2-2d-2}), and taking into account that 
$a_1+a_2=1$, one can arrive to the equality
\begin{equation} \label{eq:identity2-2d-sum}
  p\left(|e_1\rangle\right) + p\left(|e_2\rangle\right) = 2\,\mathcal{P} ,
\end{equation}
which is the desired relation between probabilities without coefficients $a_n$.

This recipe works also for arbitrary $N$. Indeed, any permutation of $N$ vectors $|e_n\rangle$ 
in Eq.~(\ref{eq:identity2}) gives rise to a valid equality; therefore one can get $N!$ equalities for 
a given set of vectors. In these $N!$ equalities, each of $N$ vectors $|e_n\rangle$ enters $(N-1)!$ times with each of $N$ 
factors $a_n$. Hence, the sum of all these equalities is
\begin{equation} \label{eq:identity2-sum}
  (N-1)! \left(\sum_{m=1}^N a_m\right) \left(\sum_{n=1}^N p\left(|e_n\rangle\right)\right) = N! \; \mathcal{P} .
\end{equation}
Finally, taking Eq.~(\ref{eq:sum-a}) into account, one can get the following relation for the function 
$p\,(|\psi\rangle)$:
\begin{equation} \label{eq:identity3}
  \forall \; \mathcal{E}: \quad
  \sum_{n=1}^N p\left(|e_n\rangle\right) = N \, \mathcal{P} .
\end{equation}

One can see now that, for $N\geq3$, the function $p\,(|\psi\rangle)$ obeys the conditions of Gleason's theorem. 
Since $p$ is a probability, it is non-negative. 
Finally, the sum~(\ref{eq:gleason-condition}) is equal to $N \, \mathcal{P}$ and therefore does not 
depend on the choice of unit vectors $|e_n\rangle$.

Thus, one can apply Gleason's theorem, that completes the proof of Eq.~(\ref{eq:answer}) for the case $N\geq3$.

\section{Case of two-dimensional state space}  \label{2d}

The above derivation of Eq.~(\ref{eq:answer}) does not cover the special case $N=2$. 
Now we will see that this case can be reduced to the case $N=4$. 

Consider a system of two non-interacting particles, each of them described by a two-dimensional state space. 
The first particle is measured by a black-box device, as shown in Fig.~\ref{fig1}. 
As above, we denote as $p\,(|\psi\rangle)$ the probability 
of flashing the light on the device, when the state of the \emph{first particle} before its measurement is $|\psi\rangle$. 
In addition, we denote as $P\,(|\Psi\rangle)$ the probability 
of flashing the light, when the joint state of the \emph{two particles} is $|\Psi\rangle$ before measurement of the first 
particle. 

Since $|\Psi\rangle$ is a vector in four-dimensional space $(N=4)$, the above derivation of Eq.~(\ref{eq:answer}) 
is valid for the function $P\,(|\Psi\rangle)$. Hence, there is such Hermitian operator $\hat A$, acting in a 
four-dimensional space and independent of $|\Psi\rangle$, that
\begin{equation} \label{eq:p-4d}
  P\,(|\Psi\rangle) = \langle \Psi | \hat A | \Psi \rangle .
\end{equation}

Let us consider the case when the first particle is in some pure state
\begin{equation} \label{eq:psi-2d}
  |\psi\rangle = \alpha |1\rangle + \beta |2\rangle ,
\end{equation}
and the second particle is in the state $|1\rangle$. (Here $|1\rangle$ and $|2\rangle$ are some basis vectors in the 
two-dimensional space.) Then, the joint state of both particles is
\begin{equation} \label{eq:psi-4d}
  |\psi\rangle |1\rangle  \equiv  \alpha |11\rangle + \beta |21\rangle .
\end{equation}
The probability of flashing the light in this situation can be expressed both as $p\,(|\psi\rangle)$ 
and as $P\,(|\psi\rangle|1\rangle)$, therefore
\begin{equation} \label{eq:p-p}
  p\,(|\psi\rangle) = P\,(|\psi\rangle|1\rangle) .
\end{equation}

Substituting Eqs.~(\ref{eq:psi-4d}) and~(\ref{eq:p-4d}) into Eq.~(\ref{eq:p-p}), 
one can express the probability $p\,(|\psi\rangle)$ as follows:
\begin{equation} \label{eq:p-4d-op}
  p\,(|\psi\rangle) = 
  \left( \alpha^* \langle11| + \beta^* \langle21| \right) \hat A \left( \alpha |11\rangle + \beta |21\rangle \right) ,
\end{equation}
i. e.
\begin{equation} \label{eq:p-4d-matrix}
  p\,(|\psi\rangle) = 
  \left( \alpha^* \; \beta^* \right) 
  \left( \begin{array}{cc} 
    \langle11|\hat A|11\rangle & \langle11|\hat A|21\rangle \\ 
    \langle21|\hat A|11\rangle & \langle21|\hat A|21\rangle 
  \end{array} \right)
  \left( \begin{array}{l} \alpha \\ \beta \end{array} \right) .
\end{equation}
The $2\times2$ matrix in the latter equation can be considered as a representation of some Hermitian operator 
$\hat A_2$, acting in the two-dimensional state space of one particle. Therefore one can rewrite 
Eq.~(\ref{eq:p-4d-matrix}) in the operator form:
\begin{equation} \label{eq:p-2d}
  p\,(|\psi\rangle) = \langle \psi | \hat A_2 | \psi \rangle ,
\end{equation}
where $\hat A_2$ does not depend on $|\psi\rangle$ (i.~e. on $\alpha$ and $\beta$). 

The derivation of Eq.~(\ref{eq:p-2d}), given in this Section, justifies Eq.~(\ref{eq:answer}) for the special case 
$N=2$, where $N$ is the dimensionality of the state space of the measured system. 
Therefore Eq.~(\ref{eq:answer}) is now proven for any measurement on any quantum system with finite $N$.

\section{From POVM to the Born rule}  \label{born-rule}

Consider some device that measures an observable $\mathcal{O}$. 
Let $p\,(|\psi\rangle)$ be the probability of getting some fixed result $\lambda$, when a 
system in a state $|\psi\rangle$ is measured by this device. 
It is already proven in Sections~\ref{thought-exp}--\ref{2d}, 
that the function $p\,(|\psi\rangle)$ can be represented in the 
form of Eq.~(\ref{eq:answer}), where $\hat A$ is some Hermitian operator. In this Section we will see that 
$\hat A$ is a projector onto an eigenspace of the operator $\hat O$, which describes the observable~$\mathcal{O}$.

Let a matrix $A_{mn}$ represents the operator $\hat A$ in a basis $|\varphi_1\rangle,\ldots,|\varphi_N\rangle$ 
of eigenvectors of $\hat O$:
\begin{equation} \label{eq:a-mn-def}
  A_{mn} \equiv \langle \varphi_m | \hat A | \varphi_n \rangle .
\end{equation}
Then, according to Eqs.~(\ref{eq:answer}) and~(\ref{eq:a-mn-def}), probabilities $p\,(|\varphi_n\rangle)$ 
are equal to diagonal matrix elements $A_{nn}$:
\begin{equation} \label{eq:p-phi-n}
  p\,(|\varphi_n\rangle)  =  \langle \varphi_n | \hat A | \varphi_n \rangle  =  A_{nn} .
\end{equation}
On the other hand, if the state of the measured system is an eigenstate of $\hat O$, then the measurement result 
must be equal to the corresponding eigenvector; therefore $p\,(|\varphi_n\rangle)$ is 1 if the $n$th 
eigenvalue is equal to $\lambda$ (i.~e. if $\hat O |\varphi_n\rangle = \lambda |\varphi_n\rangle$), and 0 otherwise. 
Hence,
\begin{equation} \label{eq:a-nn}
  A_{nn} = \left\{
  \begin{array}{l} 
    1 \quad \mbox{if } \hat O |\varphi_n\rangle = \lambda |\varphi_n\rangle, \\ 
    0 \quad \mbox{otherwise.} \\ 
  \end{array} \right.
\end{equation}

Now we will show that non-diagonal matrix elements $A_{mn}$ vanish. For this purpose, let us consider 
eigenvalues $a_1,\ldots,a_N$ of the operator $\hat A$. Since the trace of a matrix is an invariant, then  
\begin{equation} \label{eq:trace}
  \sum_n A_{nn} = \sum_k a_k .
\end{equation}
Analogously, since the sum of squared absolute values of all matrix elements is an invariant, then
\begin{equation} \label{eq:sum-squares}
  \sum_{m,n} |A_{mn}|^2 = \sum_k a_k^2 .
\end{equation}
Subtracting Eq.~(\ref{eq:trace}) from Eq.~(\ref{eq:sum-squares}), and taking into account that 
$|A_{nn}|^2 = A_{nn}$ due to Eq.~(\ref{eq:a-nn}), one can see that 
\begin{equation} \label{eq:sum-squares-nondiag}
  \sum_{m \neq n} |A_{mn}|^2 = \sum_k (a_k^2-a_k) ,
\end{equation}
where summation in the left hand side is over all non-diagonal elements.

It is easy to see that all eigenvalues $a_k$ are non-negative. Indeed, if some eigenvalue $a_k$ were negative, 
then the probability $p\,(|\chi_k\rangle)$, where $|\chi_k\rangle$ is the corresponding eigenvector, 
would be negative too:
\begin{equation}
  p\,(|\chi_k\rangle) = \langle\chi_k|\hat A|\chi_k\rangle = \langle\chi_k|a_k|\chi_k\rangle = a_k < 0 ,
\end{equation}
which is impossible. For a similar reason, $a_k$ cannot be larger than 1. Hence, all eigenvalues $a_k$ 
are bound within the range $[0,1]$ and, consequently, 
\begin{equation} \label{eq:eig-negative}
\forall n: \quad a_k^2-a_k \leq 0 .
\end{equation}
Therefore the right hand side of Eq.~(\ref{eq:sum-squares-nondiag}) is negative or zero. But the 
left hand side of Eq.~(\ref{eq:sum-squares-nondiag}) is positive or zero, so both sides are equal to zero. 
This proves that all non-diagonal matrix elements $A_{mn}$ vanish. 

So the matrix $A_{mn}$ is diagonal, and the action of the operator $\hat A$ on basis vectors $|\varphi_n\rangle$ 
is defined by Eq.~(\ref{eq:a-nn}):
\begin{equation} \label{eq:a-action}
  \hat A |\varphi_n\rangle = A_{nn} |\varphi_n\rangle = 
  \left\{
  \begin{array}{l} 
    |\varphi_n\rangle \quad \mbox{if } \hat O |\varphi_n\rangle = \lambda |\varphi_n\rangle, \\ 
    0 \qquad\; \mbox{otherwise.} \\ 
  \end{array} \right.
\end{equation}
The operator $\hat A$ is, consequently, the projector onto the eigenspace of $\hat O$ 
with eigenvalue~$\lambda$. Thus, we have seen that the postulate~(\ref{eq:projective}), 
together with the Born rule~(\ref{eq:born-rule}) in a particular case of non-degenerate eigenvalue~$\lambda$, 
are consequences of Eq.~(\ref{eq:answer}).

\section{Conclusions}  \label{conclusions}

In the main part of this paper, Sections~\ref{thought-exp}--\ref{2d}, we have presented a proof that 
probability of any result of any measurement on a quantum system, as a function of the system's 
state vector $|\psi\rangle$, obeys Eq.~(\ref{eq:answer}). (For simplicity, only systems with 
finite-dimensional state spaces were considered.) This justifies the statement that the most general 
type of measurement in quantum theory is one described by POVM.

It is important to note that this proof of Eq.~(\ref{eq:answer}) avoids using the Born rule (or any other form 
of probabilistic postulate). This opens a possibility of deriving non-circularly the Born rule from 
Eq.~(\ref{eq:answer}). Such possibility is given in Section~\ref{born-rule}. 
Note that, despite many efforts aiming to derive the Born rule (see Refs.~\cite{Dickson2011,Schlosshauer2005} 
for review), there are no generally accepted derivations up to now. Therefore the present approach may be helpful, 
due to its simplicity: all its essence is contained in three thought experiments shown in Fig.~\ref{fig4}.

Finally, let us emphasize the role of entanglement in the present derivation. 
Consideration of an \emph{entangled state of two particles} $|\Psi_N\rangle$, Eq.~(\ref{eq:psi-N}), has helped us 
to establish the probability rule for \emph{pure states of one particle alone} (not entangled with any environment).
It~seems to be that entanglement is a necessary concept for establishing the probabilistic nature 
of quantum theory.

%\section*{References}

\end{document}